\documentclass[lhep,aps,preprintnumbers,amssymb]{revtex4}

\usepackage{color}
\usepackage{epsf}

\begin{document}
\title{%
\hfill{\normalsize\vbox{%
\hbox{}
 }}\\
{Baryon asymmetry in the standard model revisited}}

\author{Renata Jora
$^{\it \bf a}$~\footnote[2]{Email:
 rjora@theory.nipne.ro}}

\affiliation{$^{\bf \it a}$ National Institute of Physics and Nuclear Engineering PO Box MG-6, Bucharest-Magurele, Romania}

\date{\today}

\begin{abstract}
We compute the baryon asymmetry in the Universe in the framework of the standard model with the only source of CP violation in the CKM matrix. Our result is within and very close to the theoretical limits for baryon asymmetry found in the literature and too small to account for the experimentally required baryon asymmetry in standard cosmologies. The method can be easily  extended to beyond standard model theories.

\end{abstract}
\maketitle

\section{Introduction}

Since the Universe is mostly composed of matter, the matter-antimatter asymmetry is an experimentally based fact that needs to be explained in all cosmological models. This asymmetry is measured by the ratio of baryon number density to photon density or entropy density ($\eta=\frac{n_b-\bar{n}_b}{n_{\gamma}}$ or $\eta=\frac{n_b-\bar{n}_b}{s}$). One of the most reliable theories for explaining baryon asymmetry is electroweak baryogenesis. This treats the generation of baryon asymmetry at the electroweak scale based on the quantum theory of particles and fields. Then the theory requires three main properties as first proposed by Sakharov \cite{Sakharov},  among which  of special significance is the presence of  CP violation.  In the standard model of elementary particles all known and verified CP violation is encapsulated in the CKM matrix.  Thus one expects that the relevant processes referring to the quark mixing terms are of maximum importance for the baryon asymmetry in the Universe.

In \cite{Kuzmin}-\cite{Farrar} the authors developed methods for computing the baryon asymmetry in the standard model.  The result in \cite{Farrar} was further disputed in \cite{Gavela}. A detailed and comprehensive analysis in the standard model was then performed in \cite{Huet}. This final estimate shows without doubt that  it is impossible in the simple context of the standard model to reach the desired baryon asymmetry requested  by the experimental data.

The calculations of the baryon asymmetry are complex and intricate and are based on certain assumptions regarding the electroweak baryogenesis. Here we will present a simple quantum field theory method that leads to results in very good agreement with the theoretical estimates in \cite{Huet} and that has direct applicability to any beyond of the standard model theory.  Section II contains the set-up for our calculation. In section III the baryon asymmetry is computed analytically in the framework of the one loop thermal effective potential of the standard model. Section IV includes the numerical results whereas section V is dedicated to Conclusions.

\section{The set-up}
The divergence of the baryon current in the standard model is given by \cite{Trodden}:
\begin{eqnarray}
\partial_{\mu}j^{\mu}_B=n_f\Bigg[\frac{g^2}{32\pi^2}W^a_{\mu\nu}\tilde{W}^{a\mu\nu}-\frac{g^{\prime 2}}{32\pi^2}F_{\mu\nu}\tilde{F}^{\mu\nu}\Bigg],
\label{divc664554}
\end{eqnarray}
where $n_f$ is the number of families, $F_{\mu\nu}$ is the electromagnetic tensor, $\tilde{F}^{\mu\nu}$ is its dual, $W^a_{\mu\nu}$ is the $SU(2)_L$ tensor and $\tilde{W}^{a\mu\nu}$ is its dual.
Here,
\begin{eqnarray}
j^{\mu}_B=\frac{1}{2}\bar{q}\gamma^{\mu}q,
\label{def64777}
\end{eqnarray}
where $q$'s are the quarks and there is an implied sum over  flavors.
Then by the Noether theorem the conserved baryonic charge is:
\begin{eqnarray}
B(t)=\int d^3 xj^0_B(x,t).
\label{bary657747}
\end{eqnarray}
The change in baryon number is then,
\begin{eqnarray}
B(t_f)-B(0)=\int_0^{t_f}dt \int d^3x\partial_{0}j_B^0=\int_0^{t_f}dt\int d^3x\partial_{\mu}j^{\mu}_B.
\label{ch64788875}
\end{eqnarray}
Since the electromagnetic field in Eq. (\ref{divc664554}) turns out irrelevant one may also write:
\begin{eqnarray}
&&\bigtriangleup B=B(t_f)-B(0)=n_f[N_{Cs}(t_f)-N_{Cs}(0)]=
\nonumber\\
&&n_f\frac{g^2}{32\pi^2}\int_0^{t_f} dt\int d^3xW^a_{\mu\nu}\tilde{W}^{a\mu\nu}.
\label{res8297546}
\end{eqnarray}
Here $N_{CS}$ is the Chern-Simons number associated to the field $W^a_{\mu}$.  Calculating the expression on the right hand side of Eq. (\ref{res8297546}) is a non-trivial task.

In what follow we will consider a simpler and equivalent approach. The quantum operator associated to the baryon divergence of current is:
\begin{eqnarray}
\langle\partial_{\mu}j^{\mu}_B\rangle=\int dA^a_{\mu}d\bar{\Psi} d\Psi dh \partial_{\mu}j^{\mu}_B\exp[i\int d^4 x {\cal L}],
\label{oprs8856}
\end{eqnarray}
where $\Psi$ and $\bar{\Psi}$ stand for all the fermion species, $A^a_{\mu}$ for all gauge bosons and $h$ for all scalars in the theory.
Eq. (\ref{oprs8856}) may be rewritten as:
\begin{eqnarray}
&&\langle\partial_{\mu}j^{\mu}_B\rangle=
\nonumber\\
&&\frac{1}{2}\int dA^a_{\mu}d\bar{\Psi} d\Psi dh[i\bar{\Psi}\frac{\partial S}{\partial \bar{\Psi}}+i\frac{\partial S}{\partial \Psi}\Psi]\exp[i\int d^4 x {\cal L}]=
\nonumber\\
&&=\frac{1}{2}48\delta(0)\int dA^a_{\mu}d\bar{\Psi} d\Psi dh \exp[i\int d^4 x [{\cal L}].
\label{res663554}
\end{eqnarray}
Here the equation of motion for the fermion fields and also the Dyson Schwinger equations were applied. Moreover we summed up over the fermions flavors. If one considers an approach in which the effective Higgs potential is determined at some relevant temperature then Eq. (\ref{ch64788875}) together with Eq. (\ref{res663554}) lead to:
\begin{eqnarray}
B(t_f)-B(0)\approx 24Vt_f\delta(0)\exp[i\Gamma(T)],
\label{fialform657775}
\end{eqnarray}
where $\Gamma(T)$ is the effective action at the temperature $T$. Here $V$ is the space volume and $t_f$ is the time interval.

The baryon asymmetry in the Universe is defined as \cite{Trodden}:
\begin{eqnarray}
\eta=\frac{n_B}{n_{\gamma}},
\label{bara5466}
\end{eqnarray}
where $n_B=n_b-\bar{n}_b$ is the difference between the baryon and antibaryons numbers per unite volume and $n_{\gamma}$ is the photon density.

One can further write Eq. (\ref{bara5466}) as \cite{Riotto}:
\begin{eqnarray}
\eta\approx\frac{\bigtriangleup_B Bn_{\gamma}}{n_{\gamma}}\approx\bigtriangleup_B B,
\label{form75664}
\end{eqnarray}
where this time $\bigtriangleup_B B=\frac{1}{2}(\bigtriangleup B-\bigtriangleup \bar{B})$.

According to our result in Eq. (\ref{fialform657775}) then:
\begin{eqnarray}
\bigtriangleup_B B=\frac{1}{2}(\bigtriangleup B-\bigtriangleup\bar{B})=12Vt_f\delta(0)[\exp[i\Gamma(T)]-\exp[i\bar{\Gamma}(T)],
\label{res775664}
\end{eqnarray}
where $\bar{\Gamma}(T)$ is the $CP$ conjugate effective action at temperature $T$.

It is more convenient to use the following definition for the baryon asymmetry:
\begin{eqnarray}
\eta_1=\frac{n_B}{s}\approx \frac{\bigtriangleup_B B}{g^*},
\label{form7756}
\end{eqnarray}
where $s\approx n_{\gamma}g^*$ is the entropy density and $g*=106.75$ is the  number of degrees of freedom in the standard model \cite{Cline}.

Our next task is to determine the difference that appears on the right hand side of Eq. (\ref{res775664}).

\section{Asymmetry of the exponential of the effective action}

The effective action of the standard model at finite temperature is obtained from the regular partition function in terms of the classical Higgs field calculated for a finite time of order $\frac{1}{T}$ \cite{Quiros}. Let us start with the definition of interest and leave aside for the moment the temperature:
\begin{eqnarray}
\exp[i\Gamma]=\int dA^a_{\mu}dhd\bar{\Psi}d\Psi\exp[i\int d^4x {\cal L}(\Phi)].
\label{effact4553}
\end{eqnarray}
Since we are actually interested in the difference between the quantity in Eq. (\ref{effact4553}) and its $CP$ conjugate we need to determine the term in the Lagrangian that is changed by $CP$ conjugation.
We will refer here only to the quark contribution. It is helpful to remove the $CP$ violating part from the quark mixing term to the down quark mass matrix (and this will not alter the results). We denote by $M_d$ the diagonal mass of the down quarks. We perform the change of variables:
\begin{eqnarray}
Ud=d'
\label{chvar466775}
\end{eqnarray}
where $U$ is the CKM matrix and $d$ represents all down quark states. Then the mass term of the down quarks becomes:
\begin{eqnarray}
\bar{d'}UM_dU^{\dagger}d'.
\label{res774663}
\end{eqnarray}
We denote:
\begin{eqnarray}
A+iB=UM_dU^{\dagger}=(U_1M_dU_1^t+U_2M_dU_2^t)+i(U_2M_dU_1^t-U_1M_dU_2^t),
\label{res666577}
\end{eqnarray}
where $A$ and $B$ are the real and imaginary parts of $UM_dU^{\dagger}$ and $U_1$ and $U_2$ are the real and imaginary parts of the $CKM$ matrix.
We separate the relevant term in the Lagrangian according to:
\begin{eqnarray}
{\cal L}={\cal L}_1-\bar{d}(A+iB)d,
\label{seo8885776}
\end{eqnarray}
for further convenience.

Then one can write Eq. (\ref{effact4553}) as:
\begin{eqnarray}
&&\exp[i\Gamma]=\int dA^a_{\mu}dhd\bar{\Psi}d\Psi\sum_{n=0}^{\infty}\Bigg[ \frac{1}{n!}[\int d^4x\bar{d}Bd]^n\Bigg]\exp[i\int d^4x[ {\cal L}(\Phi)-\bar{d}Ad]]=
\nonumber\\
&&\int dA^a_{\mu}dh d\bar{\Psi}d\Psi\sum_{n=0}^{\infty}\Bigg[ \frac{1}{n!}[iB_{ij}\frac{\partial}{\partial A_{ij}}]^n\Bigg]\exp[i\int d^4x {\cal L}(\Phi)-\bar{d}Ad]=
\nonumber\\
&&\int dA^a_{\mu}dhd\bar{\Psi}d\Psi\sum_{n=0}^{\infty}\Bigg[ \frac{(i)^n}{n!}[V_{ki}B_{ij}V_{kj}^t\frac{\partial}{\partial m_{dk}'}]^n\Bigg]\exp[i\int d^4x {\cal L}(\Phi)-\bar{d}Ad]=
\nonumber\\
&&\sum_{n=0}^{\infty}\Bigg[ \frac{(i)^n}{n!}[V_{ki}B_{ij}V_{kj}^t\frac{\partial}{\partial  m_{dk}'}]^n\Bigg]\exp[i\Gamma(A)],
\label{res55344}
\end{eqnarray}
where $\Gamma(A)$ denotes the effective action with the mass matrix $A$ for the down quarks.
Here $V$ is the unitary matrix that diagonalizes $A$ according to $VAV^t=A_d$ where $A_d$ is the matrix of eigenvalues ($m_{dk}'$)of the matrix $A$. We also used:
\begin{eqnarray}
\frac{\partial}{ \partial A_{ij}}=\frac{\partial  V_{ki}A_{ij}V_{jk}}{\partial A_{ij}}\frac{\partial}{ \partial m_{dk}'}=V_{ki}V_{jk}\frac{\partial}{ \partial m_{dk}'}.
\label{fomrytu758}
\end{eqnarray}

It is straightforward to deduce that:
\begin{eqnarray}
\exp[i\bar{\Gamma}]=\sum_{n=0}^{\infty}\Bigg[ \frac{(-i)^n}{n!}[V_{ki}B_{ij}V_{kj}^t\frac{d}{d m_{dk}'}]^n\Bigg]\exp[i\Gamma(A)].
\label{res7788677576}
\end{eqnarray}
since:
\begin{eqnarray}
{\cal L}(CP)={\cal L}_1-\bar{d}(A-iB)d.
\label{whyrtt}
\end{eqnarray}

Then Eqs. (\ref{res55344}) and (\ref{res7788677576}) lead to:
\begin{eqnarray}
&&\exp[i\Gamma]-\exp[i\bar{\Gamma}]=\sum_{n=0}^{\infty}\Bigg[ \frac{(i)^n-(-i)^n}{n!}[V_{ki}B_{ij}V_{kj}^t\frac{d}{d m_k'}]^n\Bigg]\exp[i\Gamma(A)]=
\nonumber\\
&&\sum_{n=2k+1}2\Bigg[ \frac{(i)^n}{n!}[V_{ki}B_{ij}V_{kj}^t\frac{\partial}{\partial m_{dk}'}]^n\Bigg]\exp[i\Gamma(A)].
\label{finalres6647888686}
\end{eqnarray}

\section{Baryon asymmetry in the standard model}

The one loop temperature dependent effective potential for the standard model has the expression \cite{Quiros}, \cite{Musolf}:
\begin{eqnarray}
V_{eff}(\Phi,T)=V_0(\Phi)+V_1(\Phi)+V_1^T(\Phi,T)
\label{oneloopeffect665}
\end{eqnarray}
where,
\begin{eqnarray}
&&V_0(\Phi)=-\frac{m^2}{2}\Phi^2+\frac{\lambda}{4}\Phi^4
\nonumber\\
&&V_1(\Phi)=\sum_in_i(-1)^{2s_i}\frac{1}{64\pi^2}m_i^4(\Phi)\Bigg[\ln[\frac{m_i^2(\phi)}{\mu^2}]-C_i\Bigg],
\label{epxr7746658}
\end{eqnarray}
Here i represents the particle, $s_i$ its spin, $n_i$ the number of degrees of freedom associated to the specific particle, $\mu$ is the renormalization scale and $C_i$ are constants depending on the renormalization scheme.
For example $n_g=3$ for the gauge bosons, $n_q=12$ for the quarks and $n_l=4$ for the leptons. For our calculations relevant is only $C_q=\frac{3}{2}$.

The one loop thermal corrections are given by:
\begin{eqnarray}
V_1^T=\sum_{i=boson}n_i\frac{T^4}{2\pi^2}J_b(\frac{m_i^2}{T^2})-\sum_{j=fermion}n_j\frac{T^4}{2\pi^2}J_f(\frac{m_j}{T^2}),
\label{termcorrect4566}
\end{eqnarray}
where $J_b$ and $J_b$ are loop functions with the expression:
\begin{eqnarray}
&&J_b[y^2]=\int_0^{\infty}dx x^2\ln\Bigg[1+\exp[-\sqrt{x^2+y^2}]\Bigg]
\nonumber\\
&&J_f[y^2]=\int_0^{\infty} dx x^2\ln\Bigg[1-\exp[-\sqrt{x^2+y^2}]\Bigg].
\label{func664775}
\end{eqnarray}
The following approximation for $y\ll 1$ will be useful in what follows:
\begin{eqnarray}
J_f(y^2)\approx \frac{7\pi^4}{360}-\frac{\pi^2}{24}y^2-\frac{1}{32}y^4\ln(\frac{y^2}{a_f})+O(y^3),
\label{fermfunc}
\end{eqnarray}
where $\ln(a_f)\approx2.6351$.  In practice we will use only the first two terms in the expression in Eq. (\ref{fermfunc}) as the other corrections turn out too small.
Then the effective action is:
\begin{eqnarray}
\Gamma(\Phi)=-Vt_fV_{eff}=-\frac{1}{T^4}V_{eff}.
\label{res773664}
\end{eqnarray}
Here we considered a thermal bath characterized by the volume $Vt_f=\frac{1}{T^4}$ where $T$ is the temperature and $t_f$ is the corresponding time \cite{Cline}.

We apply Eq. (\ref{res55344}) to the potential and effective action in Eqs. (\ref{oneloopeffect665})  to obtain:
\begin{eqnarray}
&&\sum_{n=0}^{\infty}\Bigg[ \frac{(i)^n}{n!}[V_{ki}B_{ij}V_{kj}^t\frac{\partial}{\partial m_{dk}'}]^n\exp[i\Gamma_{eff}(A)]\approx
\nonumber\\
&&\sum_{n=0}^{\infty}\frac{i^n}{n!}(-i)^n\Bigg[V_{ki}B_{ij}V_{kj}[-\frac{3}{4\pi^2}m_{dk}^{\prime 3}[\ln[\frac{m_{dk}^{\prime2}}{\mu^2}]-1]+\frac{1}{2}m_{dk}^{\prime}T^2]\Bigg]^n\exp[i\Gamma_{eff}(A)].
\label{res66454}
\end{eqnarray}
Here the effective potential is considered for the masses of the down quarks as extracted from the matrix A and this  does not correspond to the full down quark masses as in the standard model..

Then one can derive immediately:
\begin{eqnarray}
&&\exp[-\Gamma]-\exp[-\bar{\Gamma}]=
\nonumber\\
&&2\sum_{n=2k+1}\frac{(i)^n}{n!}\Bigg[V_{ki}B_{ij}V_{kj}(-\frac{3}{4\pi^2}m_{dk}^{\prime 3}[\ln[\frac{m_{dk}^{\prime2}}{\mu^2}-1]+\frac{1}{2}m_{d k}^{\prime}T^2]\Bigg]^n\exp[-\Gamma_{eff}(A)]=
\nonumber\\
&&2i\sin\Bigg[V_{ki}B_{ij}V_{kj}[-\frac{3}{4\pi^2}m_{dk}^{\prime 3}[\ln[\frac{m_{dk}^{\prime2}}{\mu^2}]-1]+\frac{1}{2}m_{dk}^{\prime}T^2]\Bigg]\exp[-\Gamma_{eff}(A)],
\label{res773664}
\end{eqnarray}
where this time the calculations are done in the euclidean space.

Then from Eqs. (\ref{res775664}) and (\ref{res773664}) the baryon asymmetry in the Universe reads:
\begin{eqnarray}
&&\bigtriangleup_B B=12\delta(0)Vt_f[\exp[-\Gamma (T)]-\exp[-\bar{\Gamma}(T)]=
\nonumber\\
&&i24\delta(0)Vt_f \sin\Bigg[V_{ki}B_{ij}V_{kj}(-\frac{3}{4\pi^2}[\ln[\frac{m_{dk}^{\prime2}}{\mu^2}-1]+\frac{1}{2}m_{dk}^{\prime2}T^2]\Bigg]\exp[-\Gamma_{eff}(A)]=
\nonumber\\
&&i24\delta(0)\frac{1}{T^4}\sin\Bigg[V_{ki}B_{ij}V_{kj}(-\frac{3}{4\pi^2}[\ln[\frac{m_{dk}^{\prime2}}{\mu^2}-1]+\frac{1}{2}m_{dk}^{\prime2}T^2]\Bigg]\exp[-\Gamma_{eff}(A)].
\label{finanalytres77465}
\end{eqnarray}

Here all the masses are computed at the minimum of the thermal corrected potential.
\section{Numerical calculations}

For the electroweak baryogenesis the temperature should be at the electroweak scale and it is standard to take it $T\approx100$ GeV \cite{Huet}. Moreover the perturbation theory works at best when $\mu={\rm max}[m_i(\Phi)]$ so we will take $\mu=m_t=173.3$ GeV. All the masses for the standard model fermions, Higgs or gauge bosons are taken at the electroweak scale and have the usual values indicated in \cite{PDG}. An exception are the down quarks for which the mass matrix eigenvalues are calculated according to Eqs. (\ref{res666577}). Thus in the effective action the mass $A$ is diagonalized by the matrix $V$ such that:
\begin{eqnarray}
&&VAV^t=A_d
\nonumber\\
&&A=(U_1M_dU_1^t+U_2M_dU_2^t).
\label{res69999000}
\end{eqnarray}
The matrices $U_1$, $U_2$ are the real and imaginary parts of the CKM matrix and $M_d$ is the diagonal mass matrix of the down quarks in the standard model computed at the minimum of the thermal potential. Then the eigenvalues of the down quarks that appear in the effective action are  slightly different than those in the standard model:
\begin{eqnarray}
&&m_{d1}'=z 5.42\times 10^{-3}\,\,{\rm as\,compared \,to\,} m_{d1}=4.7\times 10^{-3} \,\,{\rm GeV}
\nonumber\\
&&m_{d2}'=z 0.096\,\,{\rm as\,compared \,to\,} m_{d2}=0.096 \,\,{\rm GeV}
\nonumber\\
&&m_{d3}'=z 4.179\,\, {\rm as\,compared\, to\,} m_{d3}=4.18 \,\,{\rm GeV},
\label{res774663}
\end{eqnarray}
where $m_{dk}$ are the down quark masses in the standard model. Here  $z=\frac{\Phi_{min}}{v}$ where $\Phi_{min}$ is the minimum of the full temperature dependent effective potential.
Moreover for completeness we give here also the matrix $B$:
\begin{eqnarray}
B=(U_2M_dU_1^t-U_1M_dU_2^t).
\label{res664553}
\end{eqnarray}
where again $M_d$ is the diagonal mass of the down quarks computed at $\Phi_{\min}$.

The factor $Vt_f$ is replaced by $\frac{1}{T^4}$ where $T$ is the temperature for the thermal bath.

Next we need to regularize the $\delta(0)$ function for which we adopt the simplest procedure \cite{Srednicki}.
\begin{eqnarray}
&&\delta(0)=\lim_{x\rightarrow y}\delta(x-y)=\lim_{x\rightarrow y}\int \frac{d^4k}{(2\pi)^4}\exp[\frac{-[i\partial_{\mu}]^2}{M^2}]\exp[i(x-y)]=
\nonumber\\
&&\lim_{x\rightarrow y}\int \frac{d^4k}{(2\pi)^4}\exp[\frac{-k^2}{\mu^2}]\exp[i(x-y)]=i\frac{\mu^4}{16\pi^2},
\label{anom758844}
\end{eqnarray}
where $\mu$ is the renormalization scale and practically the highest scale in the theory besides the Higgs vacuum.

Next step is to determine the value of the effective action.  The effective potential for the standard model at the finite temperature $T$ has been calculated retaining only the relevant terms in \cite{Quiros}. It has the expression:
\begin{eqnarray}
V(\Phi_c, T)=D(T^2-T_0^2)-ET\Phi_c^3+\frac{\lambda(T)}{4}\Phi_c^4,
\label{resy8effpot366455}
\end{eqnarray}
where the coefficients are:
\begin{eqnarray}
&&D=\frac{2m_W^2+m_Z^2+2m_t^2}{8v^2}
\nonumber\\
&&E=\frac{2m_W^2+m_Z^2}{4\pi v^3}
\nonumber\\
&&T_0^2=\frac{m_h^2-8Bv^2}{4D}
\nonumber\\
&&B=\frac{3}{64\pi^2v^4}(2m_W^4+m_Z^4-4m_t^4)
\nonumber\\
&&\lambda(T)=\lambda-\frac{3}{16\pi^2v^2}\Bigg[2m_W^4\ln[\frac{m_W^2}{A_{B}T^2}]+m_Z^4\ln[\frac{m_Z^2}{A_BT^2}]-4m_t^4\ln[\frac{m_t^2}{A_FT^2}\Bigg].
\label{listcoeff455}
\end{eqnarray}
Here $\ln A_B=\ln a_b-\frac{3}{2}$ where $\ln a_b=5.4076$ and $\ln A_F=\ln a_f-\frac{3}{2}$ where $\ln a_f=2.6351$.

The minimum of the potential is compute by considering $\frac{\partial V}{\partial \Phi_C}=0$ for $T=100$ GeV. and it is given by $\Phi_{min}=208.77$ GeV.  Then the effective action at the minimum can be computed as:
\begin{eqnarray}
\Gamma_{eff}=-\frac{V_{eff}(\Phi_{min})}{T^4}.
\label{resjdhfg65}
\end{eqnarray}

As we mentioned before all the masses for the down quarks in Eq. (\ref{finanalytres77465}) must be computed at $\Phi_{min}$ such that they gain a scale factor $\frac{\Phi_{min}}{v}=0.848=z$.

Considering all the evaluations in the present section Eq. (\ref{finanalytres77465}) can be calculated and leads to:
\begin{eqnarray}
\eta_1=\frac{\bigtriangleup_B B}{g^*}=\frac{(\bigtriangleup B-\bigtriangleup\bar{B})}{2g^*}\approx -5.145\times 10 ^{-27},
\label{finalres552443}
\end{eqnarray}
where $g^*=106.75$.

\section{Conclusions}

In this work we computed the baryon asymmetry in the standard model assuming that the only source of CP violation is the CKM matrix and without considering  the exact processes that take place at the electroweak baryogenesis.

Our method is clear and straightforward and can be easily generalized to any kind of standard model extensions.  The result of the present work is in good agreement with other estimates in the literature calculated in a more complicated manner.
Thus the estimate,
\begin{eqnarray}
\eta_1=\frac{n_b-\bar{n}_b}{s}\approx -5.145 \times 10 ^{-27},
\label{finres664553}
\end{eqnarray}
is very close and within the limits calculated in \cite{Huet}:
\begin{eqnarray}
|\eta|=|\frac{n_b-\bar{n}_b}{s}|\leq6\times 10^{-27}.
\label{compr65774665}
\end{eqnarray}

  It would be of great interest to perform a similar calculation for the most common supersymmetric extensions of the standard model. This  will be done in further work.

\end{document}